\documentclass[10pt, conference, letterpaper]{IEEEtran}

\usepackage{url}
\usepackage{amssymb}
\usepackage{amsfonts}
\usepackage{amsmath}
\usepackage{amsmath,amssymb,verbatim}
\usepackage{graphicx}
\usepackage{subcaption}
\usepackage{float}
\usepackage{epsfig}
\usepackage{threeparttable}
\usepackage{algorithm}
\usepackage{algorithmic}
\usepackage{color} 
\usepackage{dblfloatfix}
\usepackage{changepage}
\usepackage{comment}
\usepackage{multirow}
\usepackage{colortbl}
\usepackage{ulem}
\usepackage{color}
\usepackage{lipsum}
\usepackage{cite}
\usepackage{tabularx}

\newcommand\blfootnote[1]{%
  \begingroup
  \renewcommand\thefootnote{}\footnote{#1}%
  \addtocounter{footnote}{-1}%
  \endgroup
}
\floatname{algorithm}{Algorithm}

\newtheorem{theorem}{\it Theorem}

\newtheorem{lemma}{\it Lemma}

\DeclareMathOperator*{\argmax}{argmax}
\begin{document}

\title{Provably Efficient Algorithms for Joint Placement and Allocation of Virtual Network Functions}

\author{
	\IEEEauthorblockN{
		Yu Sang,
		Bo Ji,
		Gagan R. Gupta, 
		Xiaojiang Du,
		and Lin Ye
	}
}

\maketitle

\blfootnote{Yu Sang, Bo Ji, and Xiaojiang Du are with the Department of Computer and
Information Sciences, Temple University, Philadelphia, PA, and Lin Ye is with Harbin Institute of Technology, China. Emails:  yu.sang@temple.edu, boji@temple.edu, gagan.gupta@iitdalumni.com, dxj@ieee.org, and hityelin@hit.edu.cn. This work was supported in part by the US NSF under grant CNS-1564128.}

\begin{abstract}
Network Function Virtualization (NFV) has the potential to significantly reduce the capital and operating expenses, shorten product release cycle, and improve service agility.
In this paper, we focus on minimizing the total number of Virtual Network Function (VNF) instances to provide a specific service (possibly at different locations) to all the flows in a network.
Certain network security and analytics applications may allow fractional processing of a flow at different nodes (corresponding to datacenters), giving an opportunity for greater optimization of resources.
Through a reduction from the set cover problem, we show that this problem is NP-hard and cannot even be approximated within a factor of $(1-o(1)) \ln{m}$ (where $m$ is the number of flows) unless P=NP.
Then, we design two simple greedy algorithms and prove that they achieve an approximation ratio of $(1-o(1))\ln{m}+2$, which is asymptotically optimal.
For special cases where each node hosts multiple VNF instances (which is typically true in practice), we also show that our greedy algorithms have a constant approximation ratio.
Further, for tree topologies we develop an optimal greedy algorithm by exploiting the inherent topological structure.
Finally, we conduct extensive numerical experiments to evaluate the performance of our proposed algorithms in various scenarios.
\end{abstract}

\section{Introduction}\label{sec:intro}
%Problem motivation
Network Function Virtualization (NFV) is emerging as a promising technology in the evolution of networking~\cite{nfv} to replace proprietary hardware appliances (e.g., middleboxes) with software modules running on general-purpose commodity servers.
These modules provide one or multiple specific network services (such as Firewalls, WAN Optimizers, Load Balancers, and Network Address Translators) called Virtual Network Functions (VNFs).
These services can be placed along the path of a network flow in a specific order (i.e., Service Function Chaining)~\cite{realtime}, potentially in a dynamic manner.
Software-Defined Networking (SDN)~\cite{sdn} is usually integrated with NFV to enable the centralized control as SDN is aimed to separate the control plane from the physical infrastructure.
This results in a highly flexible architecture and has the potential to significantly reduce the capital and operating expenses, shorten product release cycle, and improve service agility. 
In this paper, we focus on the problem of optimal placement and allocation of VNF instances to provide a specific service to all the flows in the network.

%problem statement and system model
We focus on the scenario of one single network function that requires all the data packets of the flows to be processed before they leave the network. 
This is common for many network services related to security and analytics, such as Intrusion Detection Systems (IDSs) \cite{du1,du2}, Intrusion Prevention Systems (IPSs)~\cite{nearoptimal}, Deep Packet Inspection (DPI)\footnote{In some cases, it is not required to process all the packets of a flow. For example, some DPI functions only check a small percentage of packets of a flow.
In our model, we only consider the fraction that needs to be processed.}, and network analytics/billing services. 
Each VNF instance is implemented at a virtual machine with limited resources and processing capacity. 
A network node (corresponding to a datacenter) can dynamically grow or shrink its capacity by spinning up or spinning down VNF instances.
While existing work commonly assumes that a flow is completely processed at a single node for one function (e.g.,~\cite{singleservice}), in our model we consider a more general setting where one flow may be fractionally processed at a network node and the network function can be completed at multiple nodes.
This model is based on widely adopted technologies.
For example, in the Enhanced Packet Core (EPC) of mobile networks, packets are commonly encapsulated in GTP tunnels. Packets in the same GTP tunnel can be fractionally processed at different network nodes (locations) based on the IPs of the GTP payload. Another way to do fractional processing is by computing hash functions on fields in the packet headers.

%Related work
To the best of our knowledge, existing work on VNF placement is limited to the design of heuristic algorithms, and none of the proposed algorithms can provide provable performance guarantees. 
In~\cite{realtime}, a scheduling algorithm is proposed, but this work assumes a special fat-tree topology and focuses on delay performance.
In~\cite{orchestra}, an algorithm based on dynamic programming is proposed to attack large instances of VNF placement problem.
In~\cite{singleservice}, the authors consider a model with a single type of VNF and present a heuristic algorithm towards solving the placement problem. In~\cite{stratos}, the authors propose a new architecture, called Stratos, for orchestrating VNFs outsourced to a remote cloud through traffic engineering, horizontal scaling of VNFs, etc.
In another open source project~\cite{opennf}, OpenNF is proposed to extend the centralized SDN paradigm by integrating a control plane for VNFs.
There are several other works that extend the VNF placement problem to more sophisticated applications, such as Service Function Chaining~\cite{mohammadkhan2015virtual,addis2015virtual,kuo2016deploying}. 
However, the main focus of \cite{addis2015virtual} and \cite{kuo2016deploying} is on latency and physical resource usage, respectively. For \cite{mohammadkhan2015virtual}, although a similar objective is considered, no performance guarantee is provided for their solution.
One exception that provides provable performance guarantees is the work of \cite{nearoptimal}.
However, network topology is not considered in their model.

%Solution Approach and uniqueness
We consider the problem of joint placement and allocation of VNFs (denoted by JPA-VNF) with an objective of minimizing the total number of VNF instances.
Note that even under a simplifying assumption of one single network function, the formulated problem is non-trivial (see Section~\ref{sec:nphard}). 
The main difficulty of this problem is to decide intelligently which flows, and more specifically, what fraction of the flows need to be processed at a given node so that the computing resource of the placed VNF instances is not left under-utilized.

We formulate JPA-VNF as a Mixed Integer Linear Programing (MILP) problem and prove its NP-hardness through a reduction from the set cover problem.
Using a similar reduction, we also show that JPA-VNF cannot even be approximated within a factor of $(1-o(1)) \ln{m}$ (where $m$ is the number of flows) unless P=NP.
Then, we design two simple greedy algorithms and rigorously prove that they can achieve an approximation ratio of $(1-o(1))\ln{m}+2$, which is thus asymptotically optimal.
In many NFV-based applications of practical interest, such as those in cellular networks, the flow rates are usually very large, and it typically requires multiple VNF instances at a single node (datacenter) to process the flows. 
In such scenarios, we can even prove a constant approximation ratio for our proposed greedy algorithms.
Furthermore, for networks with tree topologies, we design an optimal greedy algorithm by exploiting the inherent topological structure. 
Finally, we conduct numerical experiments both on a randomly generated dense graph and on a realistic backbone network topology of InternetMCI~\cite{internetmci}.
We evaluate the performance of our proposed algorithms in various scenarios. The simulation results show that our proposed algorithms perform very well in all the scenarios we consider.

The remainder of the paper is organized as follows. In Section~\ref{sec:model}, we describe our system model and formulate the JPA-VNF problem. In Section~\ref{sec:nphard}, we prove the hardness of the formulated JPA-VNF problem, and in Section~\ref{sec:twoalg}, we propose two simple greedy algorithms and prove that they are asymptotically optimal. Then, we consider tree topologies in Section~\ref{sec:tree} and propose an optimal algorithm. Finally, we conduct simulations in Section~\ref{sec:simulation} and make concluding remarks in Section~\ref{sec:conclusion}.

\section{System Model and Problem Formulation}\label{sec:model}

\begin{figure}[!t]
\centering
\includegraphics[trim = 0mm 0mm 0mm 0mm,clip,width=0.3\textwidth]{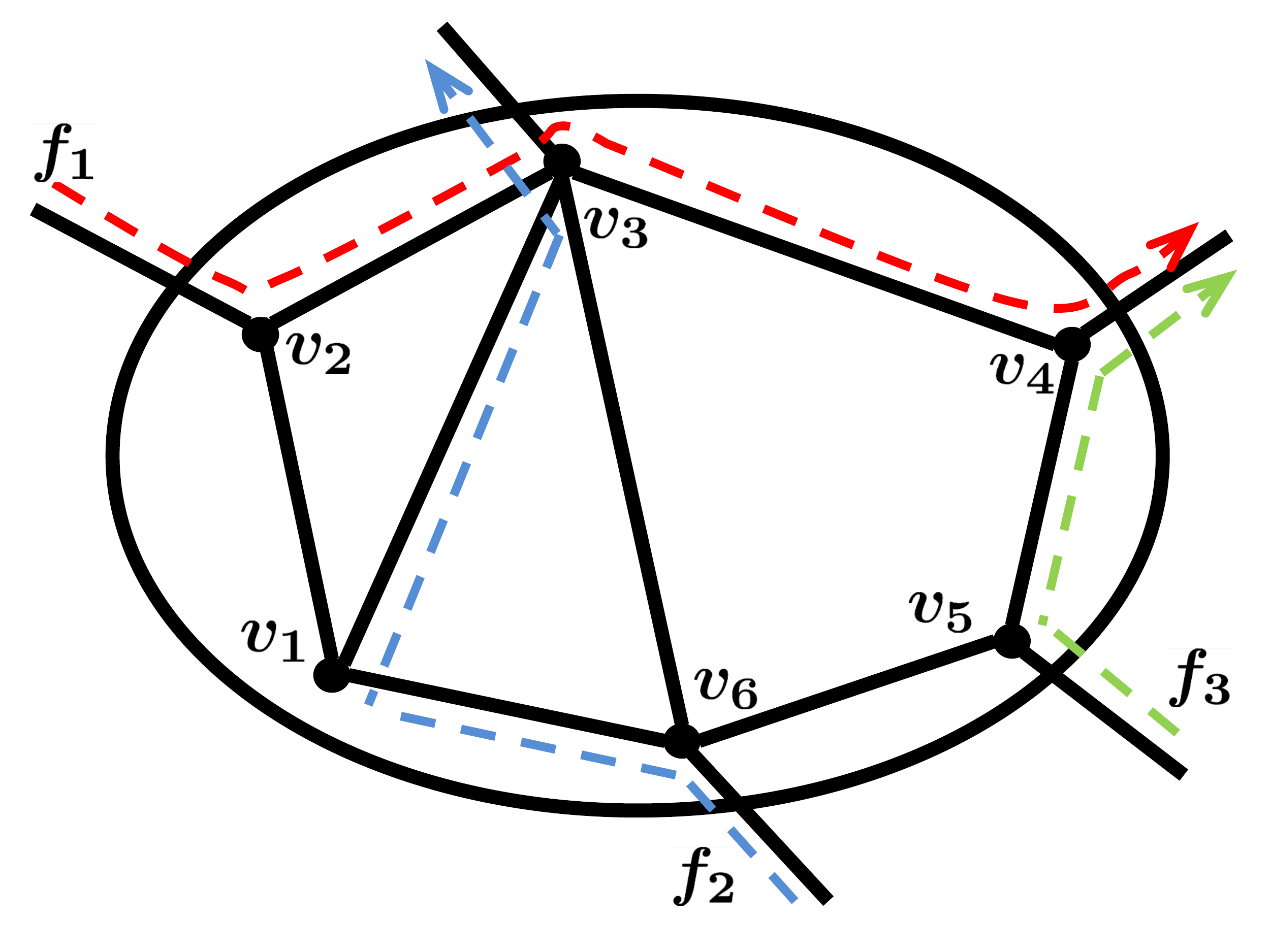}
\caption{An example network with six nodes and three flows.}
\label{general_model}
\end{figure}

We consider a network that can be modeled as a connected, undirected graph $G=(V,E)$, where $V$ is the set of nodes and $E$ is the set of edges. 
Let $n=\vert V\vert$ denote the number of nodes.
Each node represents a possible location for a VNF instance, which could be a cluster or a private datacenter owned by a certain network operator or a service provider.
Each edge denotes the link between two such locations. 
We assume that there are $m$ data flows in the network.
Let $F$ denote the set of flows.
Each data flow enters the network at a source node, traverses a sequence of links/nodes, and leaves the network after reaching the destination node. 
Let $d_j$ and $P_j$ be the flow rate and path of flow $f_j \in F$, respectively.
We assume that all the flow paths are loop free.
In particular, $P_j$ is denoted by a sequence of distinct nodes $\{v^j_1, v^j_2, \dots, v^j_{|P_j|} \}$ that are connected by a sequence of links.
Let $L_i = \{f^i_1, f^i_2, \dots, f^i_{|L_i|}\}$ denote the set of flows that pass node $v_i$.
See Fig.~\ref{general_model} for an example network.

While in the traditional networking paradigm, network functions are implemented on proprietary hardware and can only be placed at fixed locations, in a virtualized network environment VNF instances can be placed at any node as needed. 
A VNF instance is usually implemented on a standard virtual machine with limited amount of resource.
However, a node can activate multiple VNF instances to increase its total processing capacity.
In this paper, we focus on the scenario of single network function that requires all the data packets of a flow to be processed before they leave the network. 
We assume that one VNF instance has $R$ units of computing resource. 
For ease of presentation, we assume that processing one unit of data requires one unit of computing resource. 
Let $x_i$ be the number of VNF instances placed at node $v_i$. 
Then, the total processing capacity of node $v_i$ is $x_iR$. 
The resource of one node could be shared by multiple flows that pass this node.
Let $r_{ij}$ denote the amount of computing resource allocated to flow $f_j$ at node $v_i$.

Our goal is to minimize the total number of VNF instances used in the network, subject to the constraint  that all the data flows need to be fully processed before leaving the network.
This is a joint problem of placement and allocation of VNF instances: not only do we need to decide how many VNF instances to place at each node, but also need to determine how to allocate the computing resource of each VNF instance to process the flows passing each node.
We formulate JPA-VNF as the following MILP problem:
\begin{align}
\underset{x_i}{\text{min}} &~{\sum_{i=1}^n x_i}     & \notag\\
\text{subject to} &{\sum_{i:v_i\in P_j} r_{ij} \geq d_j},
				  ~\text{for all}~ 1\leq j \leq m, \label{cst:flow} \\
				  &{\sum_{j=1}^m r_{ij} \leq x_i R},
				  ~\text{for all}~ 1\leq i \leq n, \label{cst:node} \\                                
                  &~ x_i \in \{0, 1, 2 ,\dots\} . \notag
\end{align}

\setlength{\arrayrulewidth}{1pt}

We assume that a flow can be processed at multiple nodes along its path. 
Constraint~(\ref{cst:flow}) means that the total resource that flow $f_j$ receives from all the nodes on its path should be no less than its rate $d_j$.
Constraint~(\ref{cst:node}) means that the total demand for node $v_i$ cannot exceed its processing capacity.

Consider Fig.~\ref{general_model} as an example. 
There are three flows $f_1$, $f_2$, and $f_3$. 
The flow paths are $P_1=\{ v_2, v_3, v_4\}, P_2=\{v_6, v_1, v_3\}$, and $P_3=\{v_5, v_4\}$, and the flow rates are $d_1=16, d_2=6$, and $d_3=5$. 
Table~\ref{table:generaltwosolution} shows two feasible solutions, where the second solution uses three VNF instances and is optimal.

\section{Hardness of JPA-VNF}\label{sec:nphard}

In this section, we show that JPA-VNF is NP-hard through a reduction from the set cover problem, which is a well-known NP-hard problem. Similarly, we show that JPA-VNF cannot be approximated within a factor of $(1-o(1))\ln{m}$ unless P=NP. 

We start by introducing the classic set cover problem. 
Consider a set $U=\{e_1, e_2, \dots, e_m\}$ of $m$ elements  and a collection $\Phi=\{u_1, u_2, \dots, u_n \}$ of $n$ subsets of $U$.
The union of all the subsets in $\Phi$ equals $U$, i.e., $\cup^n_{i=1} u_i = U$. The objective is to find the minimum number of subsets in $\Phi$ such that their union equals $U$. 

In the following theorem, we prove the NP-hardness of JPA-VNF through a reduction from the set cover problem.\\

\begin{theorem}\label{theorem:nphard}
JPA-VNF is NP-hard.\\
\end{theorem}

\begin{table}[!t]
\centering

\begin{subtable}{\linewidth}
\centering
{
\begin{tabular}{|c| c| c| c| c| c| c|}
\cline{2-7}
\multicolumn{1}{c|}{}    & $v_1$ & $v_2$ & $v_3$ & $v_4$ & $v_5$ & $v_6$\\
\hline
 $f_1$ &   &
 \cellcolor[gray]{0.8}16 &
 \cellcolor[gray]{0.8}0 & 
 \cellcolor[gray]{0.8}0& & \\
\hline
 $f_2$ &
 \cellcolor[gray]{0.8}{6} & &
 \cellcolor[gray]{0.8}{0} & & &
 \cellcolor[gray]{0.8}{0}  \\
 \hline
 $f_3$ &
 & & &
 \cellcolor[gray]{0.8}{5} &
 \cellcolor[gray]{0.8}{0} & \\
 \hline
 \hline
 $x_i$ & 1 & 2 & & 1 & & \\
 \hline
 \end{tabular}
}
\caption{Suboptimal solution}\label{tab:1a}
\end{subtable}
\begin{subtable}{\linewidth}
\centering
{
\begin{tabular}{|c|c|c|c|c|c|c|}
\cline{2-7}
\multicolumn{1}{c|}{}
      & $v_1$ & $v_2$ & $v_3$ & $v_4$ & $v_5$ & $v_6$\\
\hline
 $f_1$ &  &
 \cellcolor[gray]{0.8}{0} &
 \cellcolor[gray]{0.8}{4} & 
 \cellcolor[gray]{0.8}{12}& &  \\
\hline
 $f_2$ &
 \cellcolor[gray]{0.8}{0} & &
 \cellcolor[gray]{0.8}{6} & & &
 \cellcolor[gray]{0.8}{0} \\
 \hline
 $f_3$ &
 & & &
 \cellcolor[gray]{0.8}{5} &
 \cellcolor[gray]{0.8}{0} & \\
 \hline
 \hline
 $x_i$ & & & 1 & 2 & & \\
 \hline
 \end{tabular}
}
\caption{Optimal solution}\label{tab:1b}
\end{subtable}
\caption{The tables above represent two solutions to the JPA-VNF problem illustrated in Fig.~\ref{general_model}.
Each row (except for the last row) corresponds to a flow.
A cell ($f_j, v_i$) is gray if flow $f_j$ passes node $v_i$. 
The processing capacity of each VNF instance is 10.
The value in gray cell $(f_j,v_i)$ is the amount of computing resource allocated to flow $f_j$ at node $v_i$.
Solution (b) requires a total of three VNF instances and is optimal.}
\label{table:generaltwosolution}
\end{table}

\begin{IEEEproof}
Given an arbitrary instance ($U,\Phi$) of the set cover problem as described above, we construct an instance ($G, F, R$) of the JPA-VNF problem. 
We show that there is a feasible solution with $k$ VNF instances for the JPA-VNF problem if and only if there exists a feasible solution of $k$ subsets in $\Phi$ for the set cover problem.

First, we set $R$ to be any positive value. 
Then, we construct a graph $G$ with $n$ nodes and a set $F$ of $m$ flows. 
For each subset $u_i \in \Phi$, we add one node $v_i$ to $G$. For each element $e_j \in U$, we construct a flow $f_j$. 
The path $P_j$ of flow $f_j$ contains $v_i$ if $e_i$ is an element of $u_j$, i.e.,  $P_j = \{v_i\vert e_i\in u_j \}$.
We create the links such that every flow can traverse the nodes in its path (e.g., making $G$ a complete graph).
We set the rate of every flow to $R/m$.
This ensures that one VNF instance is sufficient to fully process all the flows. 
Then, we calculate the set of passing flows $L_i$ for each node $v_i$. 
The objective of this JPA-VNF problem is to find the minimum number of nodes such that the union of their passing flows equals $F$.
It is easy to see that this is equivalent to finding the minimum number of subsets in $\Phi$ whose union equals the universe set $U$. 
\end{IEEEproof}

We use a simple example to illustrate the above construction process.
Consider a set cover problem with $U=\{1,2,3\}$ and $\Phi=\{u_1, u_2,\dots, u_6\}$, where $u_1=\{1,2\}$, $u_2=\{1\}$, $u_3=\{1,2\}$, $u_4=\{1,3\}$, $u_5=\{3\}$, and $u_6=\{2\}$.
Based on this set cover problem  instance and the construction in the proof  of Theorem~\ref{theorem:nphard}, the corresponding flows would be the same as those shown in Fig.~\ref{general_model}.
One difference is that the corresponding network topology is a complete graph and the flow rates are all 10/3.

In Theorem~\ref{theorem:threshold}, we state a stronger inapproximability result. The detailed proof is omitted since a reduction method similar to that in the proof of Theorem~\ref{theorem:nphard} can be applied.\\

\begin{theorem}\label{theorem:threshold}
The JPA-VNF problem cannot be approximated within a factor of $(1-o(1))\ln{m}$ unless P=NP.\\
\end{theorem}

\section{Asymptotically Optimal Greedy Algorithms}\label{sec:twoalg}

In this section, we propose two greedy algorithms and show that they can achieve an approximation ratio of $(1-o(1))\ln{m}+2$, which is thus asymptotically optimal due to the result of Theorem~\ref{theorem:threshold}.

For the set cover problem, a greedy algorithm is known to attain the best possible approximation ratio that a polynomial-time algorithm can achieve. 
It chooses a subset $u^*_i$ with  the largest number of uncovered elements in  an iterative manner until all the elements in $\Phi$ are covered. 
Inspired by this greedy algorithm for the set cover problem, we develop two greedy algorithms for JPA-VNF.
As shown in Theorem~\ref{theorem:nphard}, every subset $u_i$ in a set cover problem corresponds to a node $v_i$ in a JPA-VNF problem. 
An intuitive approach is that we treat each flow as an element in the  set cover problem and do not consider the flow rates. 
This leads to our first algorithm - the Flow Number based Greedy (\emph{FNG}) algorithm.
The FNG algorithm iteratively chooses a node with the largest number of unprocessed flows passing it. 
Another similar greedy strategy is to choose a  node that has the largest amount of unprocessed data in each iteration.
Based on this intuition, we propose our second greedy algorithm - the Flow Rate based Greedy (\emph{FRG}) algorithm. 
At first glance, FRG seems to work better since it uses additional information of flow rates. 
However, using two examples we can show that neither of them dominates.
The examples will be given at the end of this section.
More interestingly, we prove that both greedy algorithms are  asymptotically optimal.

Note that two obvious factors distinguish the JPA-VNF problem from the set cover problem. 
In the set cover problem, an element is covered as long as it is included in one of the chosen subsets. 
However, in our problem it matters which nodes are used to process a flow.
If a flow is fully processed at a node, then there is no need to allocate computing resource to this flow at the other nodes along its path. 
This resource allocation problem also leads to the second difference.
In the set cover problem, each subset has two states: either selected or not selected.
However, in JPA-VNF a node could have multiple VNF instances, which leads to a much larger state space.
Considering these two factors, when a node is selected, we choose to process all the flows that pass this node.
We design the algorithms in such a manner instead of splitting the flow rates among multiple nodes due to the following reason.
As the graph becomes larger and the flows interact with each other in a more complex way, the number of possible combinations increases exponentially. 
Our strategy is appealing because it leads to low-complexity algorithms with only minimal drop in the performance.
As we will show later, the drop in the performance is minimal. 
In addition, our proposed algorithms tend to place VNF instances at as few nodes as possible, which is preferred from the application point of view.

\begin{algorithm}[!t]
\caption{$FNG(G,F,R)$}
\begin{algorithmic}[1]\label{alg:fng}
\STATE Let $U$ be the set of all unprocessed flows. Initially, set $U=F=\{f_1, f_2,\dots, f_m\}$.
\STATE Let $S_i$ denote the set of unprocessed flows that pass node $v_i$. Initially, set $S_i=L_i$ for all $i$. \label{step:initL}
\WHILE{$\mathbf{U} \neq \emptyset$}
	\STATE Find $v_{i^*}$ such that $i^* \in {\argmax}_{i} \vert S_i\vert $. Choose the node with the smallest index $i$ when there is a tie. \label{step:pick}
	\STATE Process all the flows in $S_{i^*}$ at node $v_{i^*}$, i.e., place $x_{i^*} = \left\lceil \frac{\sum_{j:f_j\in S_{i^*}}d_j}{R}\right\rceil $ VNF instances at node $v_{i^*}$\!. \label{step:record}
	\STATE Allocate the computing resource to these unprocessed flows according to their flow rates, i.e., $r_{i^*j}=d_j$ for all $j$ such that $f_j\in S_{i^*}$.\label{step:fngprocess}
	\STATE Set $U=U\backslash S_{i^*}$.\label{step:fngupdate}
	\STATE Set $S_i=S_i\backslash S_{i^*}$ for all $i$.\label{step:fngupdate2}
\ENDWHILE
\end{algorithmic}
\end{algorithm}  

\subsection{Flow Number based Greedy Algorithm}\label{sec:fng}
We first introduce the FNG algorithm.
FNG iteratively chooses a node that covers the largest number of unprocessed flows and places just sufficient VNF instances at this node such that all the flows passing this node can be fully processed. We describe the details of FNG in Algorithm~\ref{alg:fng}.

Note that VNF instances are usually deployed on virtual machines with a limited amount of computing resource. 
In many applications like cellular networks, the flow rates can be very large, whereas there are only a limited number of datacenters in the network.
A datacenter may need a large number of VNF instances to provide certain network function or service. 
Therefore, the number of VNF instances at each datacenter is typically large, which leads to an approximation ratio close to 1. 
We first state Lemma~\ref{lemma:constant} based on this observation.
Then, we will use it to prove the main result about FNG. \\

\begin{lemma} \label{lemma:constant}
Consider the FNG algorithm. Suppose a total of $h$ VNF instances are placed at $t$ different nodes. Let $A=h/t$ be the average density of the solution. 
Suppose $A \neq 1$.
Then, FNG guarantees an approximation ratio smaller than $A/(A-1)$.\\
\end{lemma}

\begin{IEEEproof}
Let $D$ be the total amount of data rates of all the flows, i.e., $D=\sum^m_{j=1} d_j$.
Due to the way FNG functions, \emph{each node has at most one VNF instance whose computing resource is not fully used.}
Therefore, the total resource waste should be less than $tR$, i.e., $hR - D < tR$. Hence, we have
\begin{equation}
\quad D>(h-t)R = \frac{A-1}{A}hR.\label{eq:constant1}
\end{equation}

Now, we consider an optimal solution that uses a total number of $O^*$ VNF instances.
It must be satisfied that the total computing resource is no smaller than the total flow rate due to Constraint~(\ref{cst:flow}), i.e., 
\begin{equation}
O^*R\geq D .\label{eq:constant2}
\end{equation}
By combining Eqs.~(\ref{eq:constant1}) and~(\ref{eq:constant2}), we derive $h <\frac{A}{A-1}O^* $.
This implies that FNG guarantees an approximation ratio smaller than $A/(A-1)$.
\end{IEEEproof}

Lemma~\ref{lemma:constant} implies that if the average number of VNF instances at a node is greater than 1 (i.e., $A\ge 2$), the approximation ratio will be smaller than 2. 
Note that  in practice, the density of a solution could easily be much larger. 
Consider a cellular network  with 100 datacenter nodes and 10 million
flows (users) with an average flow rate of 1 Mbps. 
If the processing capacity of one VNF instance is 1 Gbps, then it is guaranteed that there will be at least 10,000 VNF instances over 100 datacenters. 
Hence, the density will be at least 100, and thus, the solution computed by FNG is guaranteed to be within $1\%$ of the optimal.
Through Lemma~\ref{lemma:constant}, we can see that processing a flow entirely at a single node not only simplifies NFV orchestration but also ensures a minimal performance loss.
Note that $A=1$ implies that there is exactly one VNF instance at every node that hosts VNF instances. 
In such cases, the JPA-VNF problem can be as difficult as the set cover problem. 

The intuition behind Lemma~\ref{lemma:constant} is as follows.
When the algorithm gives a dense solution, in which all the VNF instances are placed at a small number of nodes, rather than sparsely spreaded over the entire network, the solution is typically very close to optimal.
This is because resource waste at a single node cannot exceed $R$.
Placing the VNF instances at fewer nodes will lead to less resource waste.
We will prove the performance guarantee of FNG (stated in Theorem~\ref{theorem:order_optimal}) based on the insight  obtained from Lemma~\ref{lemma:constant}.\\

\begin{theorem}\label{theorem:order_optimal}
The approximation ratio of FNG is  no greater than $(1-o(1))\ln{m}+2$.\\
\end{theorem}

\begin{IEEEproof}
We first divide the original JPA-VNF problem into two subproblems. 
Then, we use FNG algorithm to solve these two new problems and get two solutions, a dense one and a sparse one.
We will prove that the combination of these two solutions is equivalent to the FNG solution to the original JPA-VNF problem.
The dense solution achieves a constant approximation ratio as shown in Lemma~\ref{lemma:constant}, and for the sparse one, we prove that the approximation ratio is  no greater than $(1-o(1))\ln{m}$.

Consider a JPA-VNF problem $I=(G,F,R)$. 
Assume that the total number of VNF instances used by the FNG algorithm and an optimal algorithm are $H$ and $O^*$, respectively. 
We divide the flow set $F$ into two subsets $F_1$ and $F_2$ and construct two new JPA-VNF problems. 
The partition is done is the following way.
Recall that in every iteration of the FNG algorithm, we choose one node $v_i$ and allocate VNF instances to process all the unprocessed flows that go through node $v_i$.
If we only place one VNF instance at $v_i$, then all the flows processed by this VNF instance belong to $F_1$; otherwise, they belong to $F_2$.
By doing this in each iteration, we construct flow sets $F_1$ and $F_2$.
Note that under the FNG algorithm, every flow is completely processed at a single node.
Therefore, $F_1$ and $F_2$ are disjoint, and their union equals $F$. 
We keep the original network topology and obtain two subproblems: $I_1=(G,F_1,R)$ and $I_2=(G,F_2,R)$. 
We will compare the performance of FNG and the optimal algorithm for these two subproblems.
Let $H_1$ and $H_2$ denote the number of VNF instances in the solutions given by FNG for these two subproblems, respectively, and let $O^*_1$ and $O^*_2$ denote the  number of VNF instances in the optimal solutions, respectively. 
Since $I_1$ and $I_2$ have fewer flows than $I$, we have
\begin{equation}\label{equ:optimal}
O^*_1 \le O^*\quad \text{and}\quad O^*_2 \le O^* .
\end{equation}

Now, we characterize the relation between the solutions of the new problems  (i.e., $I_1$ and $I_2$) and the original problem  (i.e., $I$) under FNG.
We will first show the following:
\begin{equation}\label{equ:sum}
H = H_1 + H_2 .
\end{equation}
Then, we will show that the following inequalities hold:
\begin{equation}\label{equ:single}
H_1 \le  (1-o(1)) O^*_1 \ln{m},
\end{equation}
and
\begin{equation}\label{equ:multi}
H_2 \le 2 O^*_2. 
\end{equation}
Finally, using these intermediate results, we compare our greedy solution with the optimal solution.

In the sequel, we want to prove Eq.~(\ref{equ:sum}). 
Consider the solutions given by FNG for problem instances $I$, $I_1$, and $I_2$.
Suppose that in the solution of $I$, node $v_i$ hosts $x_i$ VNF instances. 
We want to show the following claim: \emph{in exactly one of the two solutions of the subproblems, node $v_i$ also hosts $x_i$ VNF instances, and in the other one, node $v_i$ does not host any VNF instance.}
Eq.~(\ref{equ:sum}) follows immediately from this claim.
In order to formalize the claim, we introduce several additional notations and describe a way to compare the solution of problem $I$ and that of subproblems $I_1$ and $I_2$.
We call each while-loop iteration of problem $I$ a \emph{round}.
In each round, FNG chooses a node to place VNF instances.
Let $v_{ch}$ denote the node chosen in the $h$-th round.
If one VNF instance is placed at node $v_{ch}$ in problem $I$, then one while-loop of FNG will also be run for subproblem $I_1$, but FNG will be frozen for subproblem $I_2$.
Similarly, if more than one VNF instance is placed at node $v_{ch}$ in problem $I$, then one while-loop of FNG will also be run for subproblem $I_2$, but FNG will be frozen for subproblem $I_1$.
Let $S_i$, $S^1_i$, and $S^2_i$ denote the set of unprocessed flows at node $v_i$ in problem $I$, subproblem $I_1$, and subproblem $I_2$, respectively.
Note that $S_i$, $S^1_i$, and $S^2_{i}$ will be updated after the flows are processed in each round.

Next, we formalize the above claim and prove it by induction.
The claim consists of two parts.
\textbf{Claim: (i)} Consider the beginning of each round $h$. 
First, the following is satisfied.
\begin{equation}\label{eq:cond1}
S_i^1\subseteq S_i ~\text{and}~ S_i^2\subseteq S_i ~\text{for all}~ i,
\end{equation}
and in particular, the following holds:
\begin{equation}\label{eq:cond2}
S_{ch}^1 = S_{ch} ~\text{or}~  S_{ch}^2 =  S_{ch}.
\end{equation}
\textbf{(ii)} In all the previous $h-1$ rounds, the node chosen by FNG in problem $I$ will also be chosen by FNG in exactly one of the two subproblems $I_1$ and $I_2$; moreover, the same number of VNF instances will be placed and the same set of flows will be processed.

\textbf{Base case}: It is easy to see that at the beginning of the first round, both parts of the claim holds simply due to the way we construct the subproblems.

\textbf{Inductive step}: Suppose that the claim holds at the beginning of the $h$-th round.
We want to show that the claim also holds at the beginning of the $(h+1)$-th round.
Consider two cases based on Eq.~(\ref{eq:cond2}): i) $S^1_{ch} = S_{ch}$ and ii) $S^2_{ch} = S_{ch}$.
We first consider Case i), i.e., $S^1_{ch} = S_{ch}$.
Then, it must be the case that only one VNF instance is placed at node $v_{ch}$ in problem $I$ due to the way we construct flow set $F_1$.
Hence, in this round, one while-loop of FNG will be run for subproblem $I_1$, and FNG is frozen for subproblem $I_2$. More specifically, we will show that the same node $v_{ch}$ will be chosen in subproblem $I_1$ and a same number of VNF instances (one, in this case) will be placed.
Note that due to the greedy nature of FNG, it must be satisfied that $|S_{ch}| \ge |S_i|$ for all the nodes $v_i \in V$ since node $v_{ch}$ is chosen in problem $I$.
Then, the following must hold:
\begin{equation}\label{eq:forfrg}
\vert S_{ch}^1\vert =  \vert S_{ch}\vert \geq \vert S_i\vert \geq \vert S_i^1\vert ~\text{for all}~ i,
\end{equation}
where the first equality and the last inequality are both from the inductive hypothesis (i.e., Eq.~(\ref{eq:cond1})).
Thus, node $v_{ch}$ will also be chosen in subproblem $I_1$ and a same number of VNF instances will be placed.
This proves the second part of the claim, and it remains to show the first part.
Note that in $I$ and $I_1$, at the end of the $h$-th round, $S_i$ and $S^1_i$ will be updated: $S_i = S_i \backslash S_{ch}$ and $S^1_i = S^1_i \backslash S^1_{ch}$ for all nodes $v_i \in V$.
Thus, the relationship $S^1_i \subseteq S_i$ still holds due to $S_{ch} = S^1_{ch}$.
In subproblem $I_2$, since FNG is frozen in this round, set $S^2_i$ remains unchanged.
Since $F_1$ and $F_2$ are disjoint, we must have $S^1_{ch} \cap S^2_i = \emptyset$, and thus, $S_{ch} \cap S^2_i = \emptyset$, for all $i$.
Hence, the relationship $S^2_i \subseteq S_i$ still holds after the update of $S_i = S_i \backslash S_{ch}$.
It now remains to show Eq.~(\ref{eq:cond2}) at the beginning of the $(h+1)$-th round.
In the $(h+1)$-th round, node $v_{c(h+1)}$ will be chosen in $I$ and all the flows in $S_{c(h+1)}$ will be processed. 
If only one VNF instance is placed at node $v_{c(h+1)}$, i.e., $x_{c(h+1)} = 1$, the flows in $S_{c(h+1)}$ should be in $F_1$  due to the way we construct $F_1$.
As none of the flows in $S_{c(h+1)}$ is processed in previous rounds, they should not be processed in the subproblems, either, which means $S_{c(h+1)}\subseteq S_{c(h+1)}^1$.
We already proved $S_{c(h+1)}^1 \subseteq S_{c(h+1)}$ in the first part of the claim.
Therefore, we have $S_{c(h+1)}^1 = S_{c(h+1)}$.
On the other hand, if $x_{c(h+1)} > 1$, then we would have $S_{c(h+1)}^2 = S_{c(h+1)}$ by the same argument.
This implies that Eq.~(\ref{eq:cond2}) holds at the beginning of the $(h+1)$-th round.
For Case ii), i.e., $S_{ch}^2 = S_{ch}$, the same argument can also be applied to show that the claim still holds at the beginning of the $(h+1)$-th round. 
This completes the proof of the claim, which implies Eq.~(\ref{equ:sum}).

\textbf{Subproblem $\mathbf{I_1}$}

We now prove Eq.~(\ref{equ:single}).
We first construct another JPA-VNF problem based on $I_1$. 
Let $m_1= \vert F_1\vert $.
We change the rate of every flow in $F_1$ to a very small value.
Let $d_{min}$ be the lowest flow rate of all the flows in $F_1$.
We set the rate of all the flows to min$\{d_{min},R/m_1\}$. 
This modification ensures that every node needs at most one VNF instance to process all the flows that pass the node under any feasible algorithm, and that the rate of a flow does not increase.
We use $F_3$ to denote the set of flows with new flow rates.
Then, we construct a new JPA-VNF problem  instance $I_3=(G,F_3,R)$. 
Now, we apply FNG to $I_3$. 
Assume that FNG uses $H_3$ VNF instances to process all the flows.
FNG does not consider the rate of the flows, and every node requires only one VNF instance to process all the flows passing it.
Therefore, the solution should be the same as that for $I_1$, i.e., $H_3=H_1$. 
For this new instance, let $O^*_3$ be the optimal solution.
The only difference between $I_1$ and $I_3$ is that the later one has lower flow rates.
Therefore, it is easy to see $O^*_3\leq O^*_1$.
Note that in this case, $I_3$ can be exactly mapped to a set cover problem, and FNG also becomes equivalent to the well studied greedy algorithm for the set cover problem. 
Hence, we have
\begin{align}   
    H_1 = H_3 \notag  \le  (1-o(1))O^*_3\ln{m_1} \le  (1-o(1)) O^*_1 \ln{m},
\end{align}
where the first inequality is due to the achievable approximation ratio of the greedy algorithm for the classic set cover problem \cite{setcover}, and the second inequality is due to $O^*_3 \le O^*_1$ and $F_1 \subseteq F$. 

\textbf{Subproblem $\mathbf{I_2}$}

\begin{algorithm}[!t]
\caption{$FRG(G,F,R)$}
\begin{algorithmic}[1]\label{alg:greedy2}
\STATE Let $U$ be the set of all unprocessed flows. Initially, set $U=F=\{f_1, f_2,\dots, f_m\}$.
\STATE Let $S_i$ denote the set of unprocessed flows that pass node $v_i$. Initially, set $S_i=L_i$ for all $i$.
\WHILE{$\mathbf{U} \neq \emptyset$}
	\STATE Find $v_{i^*}$ such that $i^* \in {\argmax}_{i} \sum_{j\in S_i}d_j$. Choose the node with the smallest index $i$ when there is a tie.
	\STATE Process all the flows in $S_{i^*}$ at node $v_{i^*}$, i.e., place $x_{i^*} = \left\lceil \frac{\sum_{j:f_j\in S_{i^*}}d_j}{R}\right\rceil $ VNF instances at node $v_{i^*}$\!.\label{record2}
	\STATE Allocate the computing resource to these unprocessed flows according to their flow rates, i.e., $S_{i^*j}=d_j$ for all $j$ such that $f_j\in S_{i^*}$.
	\STATE Set $U=U\backslash S_{i^*}$.\label{update2}
	\STATE Set $S_i=S_i\backslash S_{i^*},$ for all $i$.
\ENDWHILE
\end{algorithmic}
\end{algorithm} 

Next, we prove Eq.~(\ref{equ:multi}).
The proof follows immediately from Lemma~\ref{lemma:constant}. Recall that all the nodes in $I_2$ has either no VNF instance or at least two VNF instances. 
Assume that FNG places VNF instances at $t_2$ nodes in $I_2$. Each one of these $t_2$ nodes has at least two VNF instances. 
Hence, we have $H_2/t_2 \ge 2$.
Therefore, we have $H_2\leq 2O^*_2$ from Lemma~\ref{lemma:constant}.

Combining Eqs.~(\ref{equ:optimal}), (\ref{equ:sum}), (\ref{equ:single}), and (\ref{equ:multi}),  we have
\begin{align*}
H &=H_1+H_2 \\
 & \leq (1-o(1))O^*_1\ln{m}+2O^*_2\\
 & \leq ((1-o(1))\ln{m}+2)O^*,
\end{align*} 
Therefore, the approximation ratio of FNG is upper bounded by $(1-o(1))\ln{m}+2$.
This, along with Theorem~\ref{theorem:threshold}, implies that FNG is asymptotically optimal.
\end{IEEEproof}

\subsection{Flow Rate based Greedy Algorithm}

We now introduce the FRG algorithm. 
Similar to FNG, FRG iteratively chooses a node with the largest unprocessed flow rate.
The details of FRG are provided in Algorithm~\ref{alg:greedy2}.

As we mentioned earlier, even though the FRG algorithm considers the flow rate information when making decisions, it does not necessarily  guarantee a better performance than FNG.
In Table~\ref{table:twoinstances}, we provide two simple examples to show that neither of the greedy algorithms dominates in general.
In the following, we show that FRG also achieves an approximation ratio of $(1-o(1))\ln{m}+2$.\\

\begin{theorem}\label{theo:frg}
The approximation ratio of FRG is  no greater than $(1-o(1))\ln{m}+2$.\\
\end{theorem}

\begin{IEEEproof}
The proof is similar to that of Theorem~\ref{theorem:order_optimal}.
We can also divide the original problem into two subproblems as described in the proof Theorem~\ref{theorem:order_optimal}.
Every subproblem consists of the original topology and a subset of flows.
The only difference is in the proof of Eq.~(\ref{equ:sum}). 
Eq.~(\ref{equ:sum}) shows that if we run the algorithm on these two subproblems, we will still place the same number of VNF instances at the nodes as the original problem.
The basic idea is that the partition will not affect which nodes we choose in every round.
The proof is similar to that for FNG. 
Details are omitted here.
The key is to show that Eq.~(\ref{eq:cond1}) holds at the beginning of  every round. 
At the beginning of the $h$-th round, Eq.~(\ref{eq:forfrg}) should be rewritten as
\[
\sum_{j\in S_{c,h}^1} d_j   =    \sum_{j\in S_{c,h}} d_j   \geq  \sum_{j\in S_i} d_j   \geq  \sum_{j\in S_i^1} d_j  ~\text{for all}~ i.
\]
Therefore, Eq.~(\ref{equ:sum}) still holds. 
It is easy to see that Eq.~(\ref{equ:single}) also holds because when we set all the flow rates to be the same, FRG is the same as FNG. Eq.~(\ref{equ:multi}) holds since it does not involve the flow rate. 
Combining Eqs.~(\ref{equ:sum}), (\ref{equ:single}), and (\ref{equ:multi}) implies that FRG can achieve the same approximation ratio as FNG.
\end{IEEEproof}

\begin{table}[!t]
\centering
\begin{subtable}{.45\linewidth}
\centering
{
\begin{tabular}{|c|c|c|c|}
\cline{2-4}
\multicolumn{1}{c|}{} & $v_1$ & $v_2$ &flow rate\\
\hline
 $f_1$ & \cellcolor[gray]{0.8}{}  &  & 10 \\
\hline
 $f_2$ & \cellcolor[gray]{0.8}{}   
 	   &  & 10 \\
 \hline
 $f_3$ & \cellcolor[gray]{0.8}{}   
 	   & \cellcolor[gray]{0.8}{} 
 	   & 4 \\
 \hline
  $f_4$ &  & \cellcolor[gray]{0.8}{} & 26\\
 \hline
 \end{tabular}
}
\caption*{FRG: ($x_1=2,x_2=3$);\\ FNG: ($x_1=3,x_2=3$).}
\caption{}
\end{subtable}
\begin{subtable}{.45\linewidth}
\centering
{
\begin{tabular}{|c|c|c|c|}
\cline{2-4}
\multicolumn{1}{c|}{} & $v_1$ & $v_2$ &flow rate\\
\hline
 $f_1$ & \cellcolor[gray]{0.8}{}   
 	   &  & 3 \\
\hline
 $f_2$ & \cellcolor[gray]{0.8}{}
 	   &  & 1 \\
 \hline
 $f_3$ & \cellcolor[gray]{0.8}{}   
 	   & \cellcolor[gray]{0.8}{}   
 	   & 6 \\
 \hline
  $f_4$ &  & \cellcolor[gray]{0.8}{} 
   & 10\\
 \hline
 \end{tabular}
}
\caption*{FRG: ($x_1=1,x_2=2$);\\  FNG: ($x_1=1,x_2=1$).}
\caption{}
\end{subtable}
\caption{Consider a very simple graph with only two nodes and an edge connecting them. Let $R=10$. 
While the FRG algorithm is better in instance (a), the FNG algorithm is better in instance (b). }
\label{table:twoinstances}
\end{table}

\subsection{Complexity Analysis}
Next, we analyze the complexity of FNG.
In line~\ref{step:initL}, we need to calculate $L_i$, which is the set of flows passing node $v_i$.
To do this, we need to go through all the paths $P_j$. 
The maximum path length is upper bounded by $n$ since we assume loop-free paths. 
Hence, the running time of this step is $O(mn)$. 
We also need to keep an unprocessed flow set $S_i$ for every node.
Initially, $S_i = L_i$.
In each while loop, we need to find a node $v_{i^*}$ with the largest $\vert S_{i^*}\vert$, process all the flows in $S_{i^*}$, and update $S_i$ for all $i$. 
One simple implementation is to update all $S_i$ and to get the largest $\vert S_i\vert$ at the end of the update in each while loop. 
Since there are at most $n$ while loops, and in each while loop, there are at most $n$ comparisons, the running time of this step is $O(n^2)$. 
Note that here we have not computed the cost for updating $S_i$ yet.
It is easy to see that the total cost for the updating process in the while loops is the same as that in line~\ref{step:initL} where we initialize $L_i$ for all $i$. 
Hence, the total running time for the while loops is $O(n^2 + mn)$, and thus, the overall running time of FNG is also $O(n^2 + mn)$. 
We can analyze the complexity of FRG in a similar manner and show that the complexity of FRG is also the same as that of FNG.

\section{An Optimal Algorithm for Tree Topology}\label{sec:tree}
\begin{algorithm}[!t]
\caption{$GFT(T,F,R)$}
        \begin{algorithmic}[1]\label{alg:tree}
        	\FOR{$p$ from the largest to the smallest}
        	 	\FOR{$q=1 \to l_p $}
        		    \IF {there are flows leaving the network through $v_{p,q}$}
                        \STATE put $\lceil d_{p,q}/R\rceil$ VNF instances at $v_{p,q}$ to process all the flows leaving the network through $v_{p,q}.$\label{alg:process1}
                    	\WHILE{there is computing resource left}
                    		\STATE Allocate the computing resource to process the first flow $f_j$ in $F_{p,q}$. \label{alg:process2}
                    		\STATE Update the rate of $f_j$ if it is not fully processed. Otherwise, move it out of $F_{p,q}$. 
                    \ENDWHILE
                    \STATE Let $F'$ be the set of flows that are fully processed in this loop (line 3 $\sim$ line 8). Update the waiting list and unprocessed flow set of all other nodes, $F_{s,k}= F_{s,k}/F'$, $D_{s,k}= D_{s,k}/F'$ for all $s$ and $k$.
                    \ENDIF 
        		\ENDFOR
        	\ENDFOR
        \end{algorithmic}
\end{algorithm}

As described in our general model, for network operators who have their own datacenters within the core network, they may choose to implement their VNF instances that are scattered over different locations~\cite{cloud4nfv}\cite{cloud4nfv2}. 
This general model leads to an NP-hard problem as we described in Section~\ref{sec:nphard}.
However, some network services may require the network to have special topologies. 
Tree topologies are widely used for streaming services and Content Delivery Networks (CDNs)~\cite{cdn}.
In such cases, by harnessing the properties of tree topologies, we propose an optimal solution for JPA-VNF under some simplifying assumptions.

\subsection{An Optimal Algorithm for JPA-VNF with Tree Topology}

We consider a tree  network topology, denoted by $T$. 
Let $l_p$ denote the number of nodes at the $p$-th level of the tree.
We assume that the root is at level 1, which is the highest level.
From left to right, all these $l_p$ nodes are denoted by $\{ v_{p,1}, v_{p,2},\dots ,v_{p,l_p}\!\}$.
We assume that all the flows are upstream flows (i.e., from a lower-level node to a higher-level node in the tree).
We make this assumption for ease of presentation only; our results can be immediately generalized to cases where the flows are either upstream or downstream.
Let $T_{p,q}$ be the subtree rooted at node $v_{p,q}$.

\begin{figure}[!t]
\centering
\includegraphics[trim = 0mm 0mm 0mm 0mm,clip,width=0.4\textwidth]{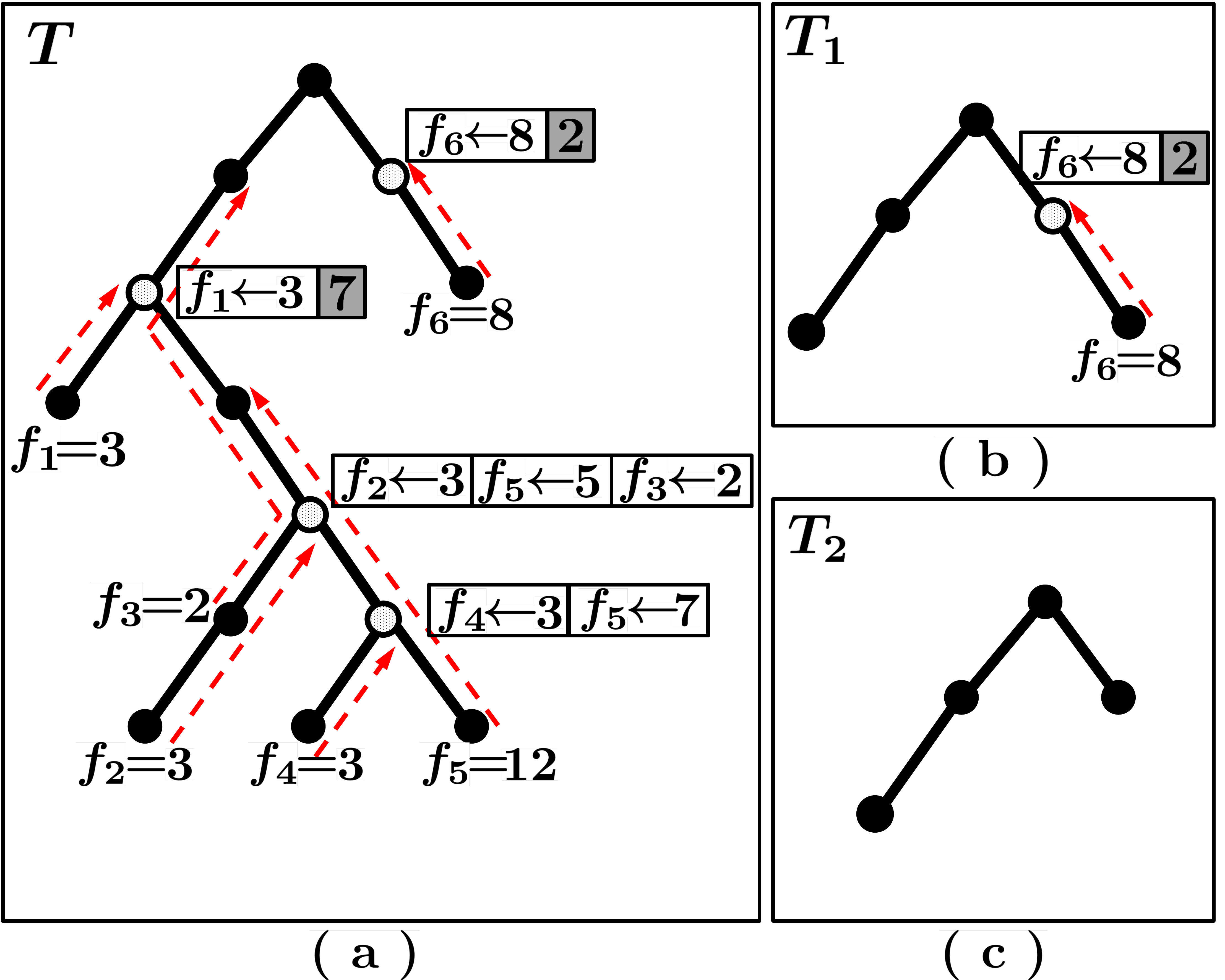}
\caption{An optimal solution generated by GFT for JPA-NFV with tree topology. Dashed lines denote the flow paths. A node is denoted by a hollow cycle if there is at least one VNF instance placed at the node. The rectangles next to the hollow nodes present the allocation of computing resource to the flows. Assume that the computing capacity of each VNF instance is $R=10$.}
\label{fig:tree1}
\end{figure}

Our algorithm is based on a key observation: if we check all the nodes on the path of a flow in a bottom-up manner, we should not process the flow until it intersects other flows or it is about to leave the network. 
This is because processing a flow at a lower-level node may lose the opportunity to combine it with other flows at a higher level. 
Hence, a good strategy would be to not process the flow until it reaches the highest-level node along its path (i.e., at the node through which the flow leaves the network).
Now, we propose our greedy algorithm based on this key idea. 
We call this algorithm Greedy For Tree  \emph{(GFT)}, which traverses all the nodes in the tree from the lowest level to the root node. 
Let $D_{p,q}$ be the set of all unprocessed flows leaving the network through node $v_{p,q}$ and $d_{p,q}$ be the total rate of all the flows in $D_{p,q}$.
Once we reach a node $v_{p,q}$ through which a flow leaves the network, we place VNF instances at this node to process all the flows in $D_{p,q}$.
Then, the problem would be how to allocate the remaining computing resource if the flow does not consume all the resource. 
For every node $v_{p,q}$, we create a waiting list $F_{p,q}$, which consists of all the unprocessed flows going through $v_{p,q}$. 
These flows are sorted in a nonincreasing order of  the level of the node through which a flow leaves the network.
Then, we allocate all the available computing resource to the first flow in the waiting list.
The detailed operations of this algorithm are provided in Algorithm~\ref{alg:tree}.
To help understand the operations of GFT, we provide an example in Fig.~\ref{fig:tree1}a and present the detailed steps of this example in Table~\ref{table:allocation}.

\subsection{Main Result: Optimality}\label{sec:proofofopt}

In Theorem~\ref{theo:treeopt}, we state our main result of optimality.\\

\begin{theorem}\label{theo:treeopt}
GFT is optimal for tree topologies.\\
\end{theorem}

A key insight from our investigations for general graphs is to minimize the waste of resource by processing multiple flows together at the same node.
Therefore, in order to prove the optimality of GFT, we need to check the nodes where resource waste happens.
We define such nodes as \emph{Breaking Points}.
Consider a JPA-VNF problem and a feasible solution for this problem.
A node is called a \emph{Breaking Point} if it hosts a VNF instance whose computing resource is not fully utilized.
Breaking points have a very important property in one particular type of solutions, which we define as \emph{conservative solutions}. 
A solution is called conservative if every breaking point in the solution hosts at most one VNF instance that is not fully utilized.
Apparently, the solution given by GFT is conservative.
We further introduce another notion called \emph{external flows} and then state the property of breaking points in Lemma~\ref{lemma:simpcase}, which will be used to prove Theorem~\ref{theo:treeopt}.
For a node $v_{p,q}$, if the path of a flow has exactly one end within the subtree $T_{p,q}$ (including $v_{p,q}$), we call this flow an \emph{external flow} of node $v_{p,q}$.\\

\begin{lemma}\label{lemma:simpcase}
Consider a conservative solution for a JPA-VNF problem with tree topology. 
Let $v_{p,q}$ be a breaking point. 
Suppose that node $v_{p,q}$ is the only breaking point within subtree $T_{p,q}$ and that $v_{p,q}$ does not have any external flows. 
Then, no other feasible algorithm can use fewer VNF instances in $T_{p,q}$ than the conservative solution.\\
\end{lemma}

\begin{table}[!t]
\centering
\begin{tabular}{|c|c|c|c|c|c|}
\hline
\multirow{2}{0.3cm}{Step} & 
\multirow{2}{0.4cm}{Node} & 
\multirow{2}{1cm}{Leaving flow} & 
\multirow{2}{1cm}{$F_{p,q}$} &
\multirow{2}{1.2cm}{\# of VNFs} &
\multirow{2}{1cm}{Resource allocation}
\\
 & & & & & \\
 \hline
\multirow{2}{0.3cm}{1} & 
\multirow{2}{0.4cm}{$v_{6,2}$ } & 
\multirow{2}{0.3cm}{$f_4$ } & 
\multirow{2}{1cm}{$f_4=3$ $f_5=12$} &
\multirow{2}{1.2cm}{$\lceil \frac{3}{10}\rceil =1 $ } &
\multirow{2}{1cm}{$3\rightarrow f_4$ $7\rightarrow f_5$}
\\
 & & & & & \\ 
 \hline
\multirow{3}{0.3cm}{2} & 
\multirow{3}{0.4cm}{$v_{5,1}$ } & 
\multirow{3}{0.3cm}{$f_2$ } & 
 \multirow{3}{1cm}{$f_2=3$ $f_5=5$ $f_3=2$} &
\multirow{3}{1.2cm}{$\lceil \frac{3}{10}\rceil =1 $ }&
\multirow{3}{1cm}{$3\rightarrow f_2$ $5\rightarrow f_5$ $2\rightarrow f_3$}
\\
 & & & & & \\ 
 & & & & & \\ 
  \hline
\multirow{2}{0.3cm}{3} & 
\multirow{2}{0.4cm}{$v_{3,1}$ } & 
\multirow{2}{0.3cm}{$f_1$ } & 
 \multirow{2}{1cm}{$f_1=3$}&
\multirow{2}{1.2cm}{$\lceil \frac{3}{10}\rceil =1 $ } &
\multirow{2}{1cm}{$3\rightarrow f_1$}
\\
 & & & & & \\ 
  \hline
\multirow{2}{0.3cm}{4} & 
\multirow{2}{0.4cm}{$v_{2,2}$ } & 
\multirow{2}{0.3cm}{$f_6$ } & 
\multirow{2}{1cm}{$f_6=8$}&
\multirow{2}{1.2cm}{$\lceil \frac{8}{10}\rceil =1 $ }  &
\multirow{2}{1cm}{$8\rightarrow f_6$}
\\
 & & & & & \\ 
  \hline

\end{tabular}
\caption{This table shows how to allocate VNF instances to the network shown in Fig.~\ref{fig:tree1} a under GFT.}
\label{table:allocation}
\end{table}
\begin{IEEEproof}
Let $N_T$ be the number of VNF instances in $T_{p,q}$ under the conservative solution.
Note that $T_{p,q}$ only has one breaking point which is the root node.
In this case, the VNF instances within $T_{p,q}$ only process the flows whose full paths are within the subtree.
Let $F_t$ be the set of all such flows. 
Note that all the flows in $F_t$ must be processed within $T_{p,q}$.
The total rate of the flows in $F_t$ is $\sum_{f_j\in F_t} d_j$.
Since the solution is conservative, computing resource waste occurs for at most one VNF, which must be at node $v_{p,q}$.
This implies $\sum_{f_j\in F_t} d_j >(N_T-1)R$.
Therefore, no other  feasible algorithm can place fewer than $N_T$ VNF instances in $T_{p,q}$ to process all the flows in $F_t$.
\end{IEEEproof}

The key idea of the proof is as follows.
If none of the breaking points have external flows, we can iteratively remove the subtrees rooted at breaking points in a bottom-up manner.
In each iteration, we remove a subtree with only one breaking point, which is its root.
The solution given by GFT is conservative.
Lemma~\ref{lemma:simpcase} implies that no  feasible algorithm can use fewer VNF instances in the subtrees.
Every time we remove such a subtree, we construct a new JPA-VNF problem instance based on the remaining topology.
Since there is no external flow for the subtree's root (i.e., the breaking point), the VNF instances left in the network can still form a conservative solution for the new instance.
By doing this repeatedly, we can show that the algorithm achieves optimality after all breaking points are removed.
However, breaking points can have external flows. 
In the following, we show that  processing these external flows actually do not increase the number of VNF instances.
We can simply remove all the external flows of the breaking points iteratively and get the simpler case as described above.
The detailed proof is provided in the following.\\

\begin{IEEEproof}[Proof of Theorem~\ref{theo:treeopt}] Consider a JPA-VNF problem $I = (T,F,R)$ on a tree topology $T$.
Let $N_g$ and $N_o$ be the number of VNF instances used by GFT and an optimal algorithm.
Our goal is to show the following:
\begin{equation} \label{eq:ngo}
N_g\leq N_o.
\end{equation}
We first remove all the external flows of the breaking points such that none of the breaking points have external flows.
After removing external flows, we do not change the placement of VNF instances. 
In this case, the computing resource allocated to process these flows will be wasted and may create new breaking points.
To assist the analysis, we create a priority queue that consists of all the breaking points.
The breaking points are sorted in a nondecreasing order of the level.
The breaking points at the same level are sorted from left to right to break the tie.
In each iteration, we check the breaking point at the head of the queue.
We remove all the external flows of this breaking point if there is any and then remove it from the queue.
If there are new breaking points generated in this process, those new breaking points are inserted into the priority queue.
We repeat the procedure until the priority queue becomes empty.
Note that the external flows of a breaking point must already be fully processed within the subtree rooted at this breaking point.
Otherwise, the remaining resource of this breaking point would have been allocated to process it.
Therefore, new breaking points would only appear at a lower level. 
Hence, this procedure scans all the breaking points (including the new ones coming up during this procedure) of the tree from the root to the leaves.

Next, we want to show that throughout the above procedure, removing external flows of a breaking point does not reduce the number of VNF instances placed within the subtree rooted at each breaking point.
We prove this by contradiction. 
Suppose that the number of VNF instances decreases after an external flow is removed. 
Then, there must exist at least one VNF instance, whose computing resource is entirely used to process external flows.
However, this could not happen because GFT would not activate a new VNF instance for an external flow in the first place.
This implies for all the nodes that host VNF instances, only part of the computing resource of one VNF instance is used to process external flows of this node.
This property also ensures that each of the new breaking points  hosts at most one VNF instance that is not fully utilized.

We now consider the system with all the external flows removed for each breaking point.
Let $F_r$ denote the set of remaining flows.
We can construct a new JPA-VNF problem $I' = (T,F_r,R)$.
Assume that there are $k$ breaking points in $I^{\prime}$, including all the new breaking points.
We denote the set of all breaking points by $V^{\prime}= \{ v_{p_1,q_1}\!, v_{p_2,q_2}\!, \cdots\!, v_{p_k,q_k}\!\}$. 
The breaking points are sorted in a nondecreasing order of their level, i.e., $p_1\geq p_2 \geq \cdots \geq p_k$. 
The nodes at the same level are sorted according to the second index $q$. 
Note that there remain $N_g$ VNF instances in the system.
As mentioned earlier, every breaking point hosts at most  one VNF instance that is not fully utilized.
Therefore, these $N_g$ VNF instances form a conservative solution for $I'\!$.
Assume that an optimal solution uses $N'_o$ VNF instances to solve $I'\!$. 
It is easy to see that $N^{\prime}_o \le N_o$ since $F_r \subseteq F$.
Therefore, in order to prove Eq.~(\ref{eq:ngo}), it is sufficient to show the following: 
\begin{equation}\label{eq:obj}
N_g\leq N'_o.
\end{equation}

In the sequel, we prove Eq.~(\ref{eq:obj}). 
We iteratively  remove the subtrees rooted at the breaking points, by starting with the breaking point at the lowest level (i.e., $v_{p_1,q_1}\!$) and the corresponding subtree $T_{p_1,q_1}\!$.
According to Lemma~\ref{lemma:simpcase}, no algorithm can put fewer VNF instances in $T_{p_1,q_1}\!$.
Let $N_{gs}^1$ and $N_{os}^1$ be the number of VNF instances in subtree $T_{p_1,q_1}\!$ for our solution and the optimal solution, respectively. 
Then, the following inequality follows from Lemma~\ref{lemma:simpcase}:
\begin{equation}\label{eq:it1}
N_{gs}^1 \leq N_{os}^1.
\end{equation}
We remove subtree $T_{p_1,q_1}\!$ from $T$ and also remove all the flows within $T_{p_1,q_1}\!$.
Let $T^1$ be the remaining topology.
Let the set of remaining flows be $F_r^1\subseteq F_r$.
Now, we have a new JPA-VNF problem $I^1 = (T^1, F_r^1,R)$.

We use $N_g^1$ and $N_o^1$ to denote the number of VNF instances left on $T^1$ after removing  $T_{p_1,q_1}\!$ under our algorithm and the optimal algorithm, respectively.
Note that $N_g^1 = N_g - N_{gs}^1$ and $N_o^1 = N^{\prime}_o - N_{os}^1$.
Due to Eq.~(\ref{eq:it1}), in order to show Eq.~(\ref{eq:obj}), it remains to show  $N_g^1 \leq N_o^1$. 

We repeat the above procedure and argument until all the $k$ breaking points are removed. 
Then, there are two cases for the remaining topology: (i) it is empty; and (ii) it is a tree without any breaking point. 
Case (i) is trivial. 
In Case (ii), let $N^k_g$ and $N^k_o$ denote the number of VNF instances left in the remaining topology.
Since there is no breaking point, there is no resource waste for the $N^k_g$ VNF instances. 
Hence, we have $N^k_g \le N^k_o$.
This completes the proof.
\end{IEEEproof}

\begin{figure*}[t!]
        \centering
                \includegraphics[width=\textwidth]{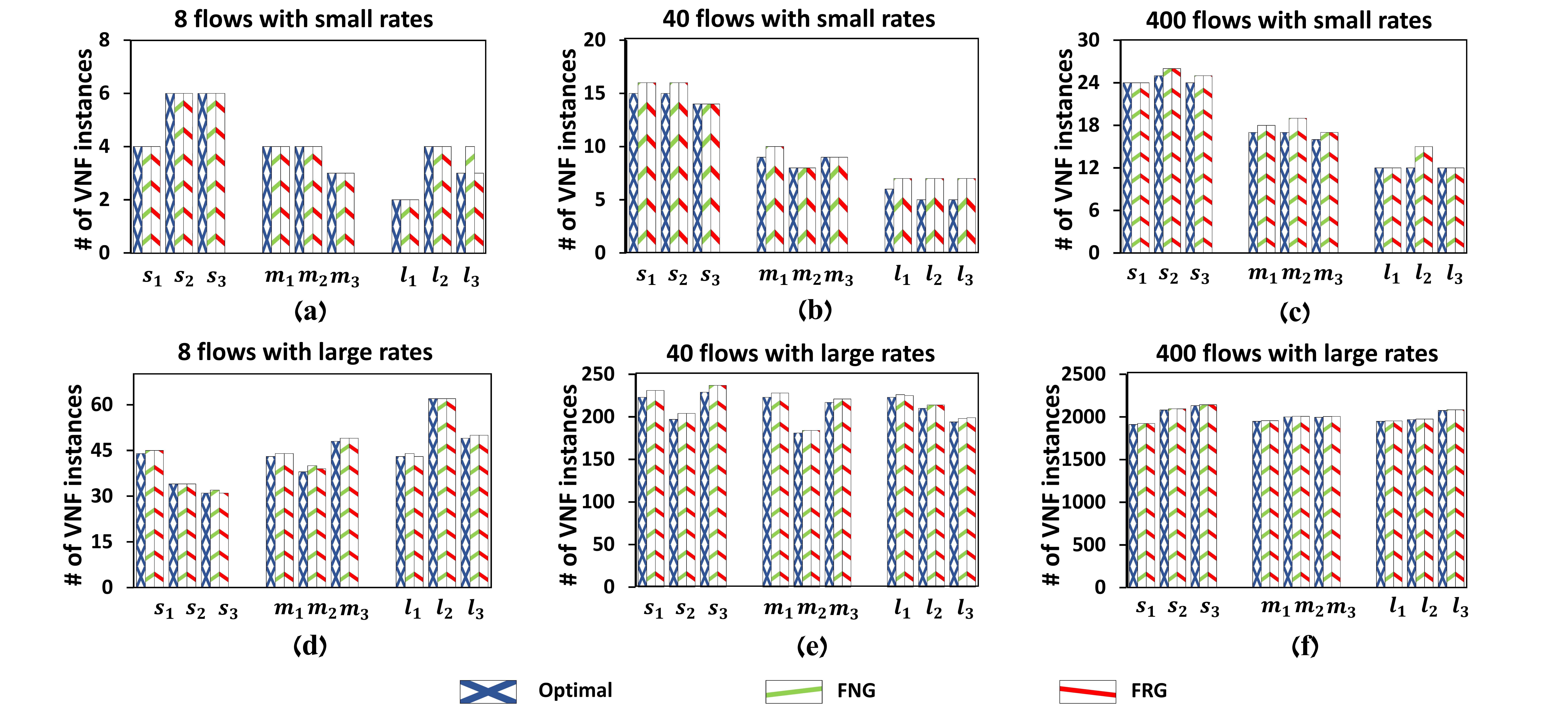}
				\caption{Simulation results for the random topology.}
				\label{fig:simulation40}
\end{figure*}

Next, we analyze the complexity of GFT.
The first step is the same as FNG, which is to go through all $P_j$ to build $L_i$.
One difference is that we also need to record which flows leave the network through node $v_i$.
However, this does not affect the running time.
In our model, the longest path in the tree topology is from the root to the leaf node at the lowest level.
Therefore, the maximum path should be $\log{n}$. 
Similar to the analysis for FNG algorithm,  the complexity of the first step for GFT is $O(m\log{n})$.
In the second step, we go through all the nodes. 
We allocate resources to a node when there are flows leaving the network through this node.
Once the flows passing the node are processed, we need to notify the nodes on the paths of these flows to update their $L_i$.
The algorithm ends after we go through all the nodes. 
Then, if we do not consider the updating process, the running time should just be $O(n)$. 
In total, the updating process will cost $O(m\log{n})$ running time because the updating process costs the
same time as that for building $L_i$ in line 1. 
Therefore, the second step costs $O(n+m\log{n})$ and the time complexity of GFT is $O(n + m\log{n})$.

\begin{figure}[!t]
		\centering
        \includegraphics[width=0.35\textwidth]{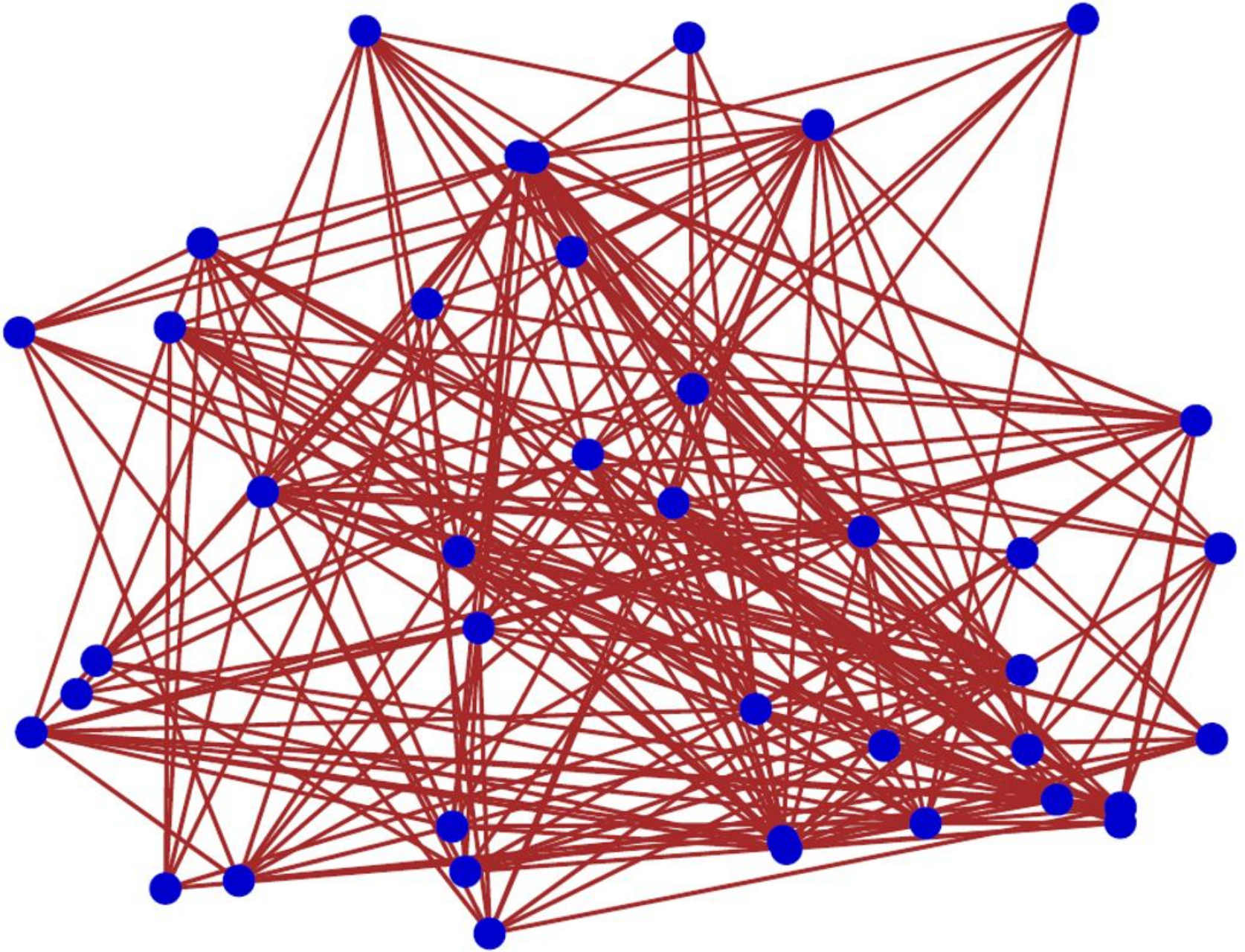}
        \caption{A randomly generated topology with 40 nodes.}
        \label{fig:randomgraph}
\end{figure}

\section{Numerical Results}\label{sec:simulation}

\begin{figure}[t!]
        \centering
                \includegraphics[width=0.45\textwidth]{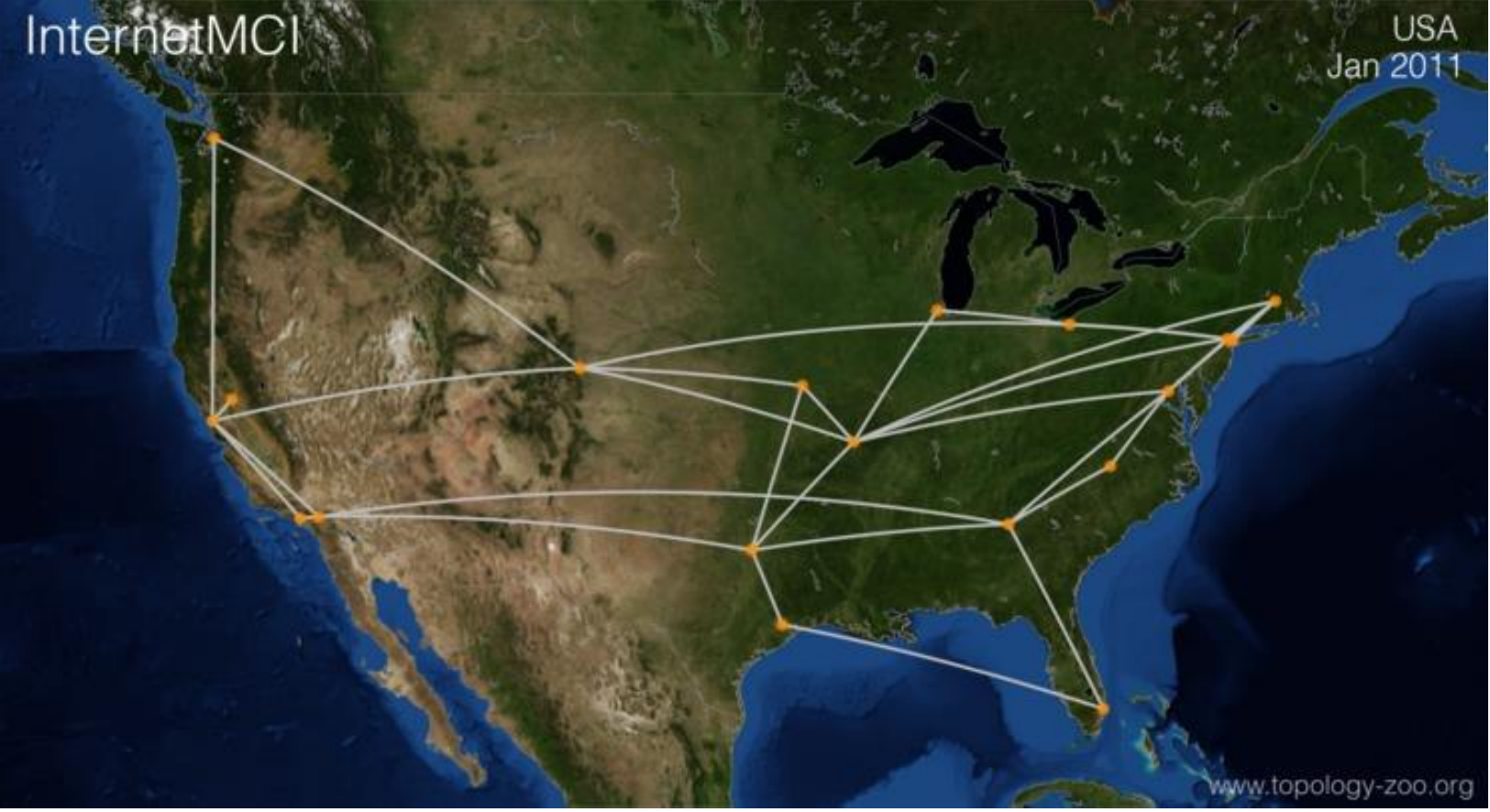}
				\caption{Topology of InternetMCI.}
				\label{fig:internetmci}
\end{figure}

In this section, we evaluate the performance of our proposed greedy algorithms in various scenarios. 
We conduct simulations both for a randomly generated network topology (Fig.~\ref{fig:randomgraph}) and for a realistic backbone network topology of InternetMCI   (Fig.~\ref{fig:internetmci})  \cite{internetmci}.

The randomly generated network topology consists of 40 nodes and 234 links. We assume that each VNF instance has a processing capacity of 10, i.e., $R=10$. 
We evaluate the performance of our proposed greedy algorithms by comparing them with the optimal solution computed by the GNU Linear Programming Kit (GLPK)~\cite{glpk}.
For the problem instances we consider, GLPK computes an optimal solution within a reasonable amount of time.

\begin{figure*}[t!]
        \centering
                \includegraphics[width=\textwidth]{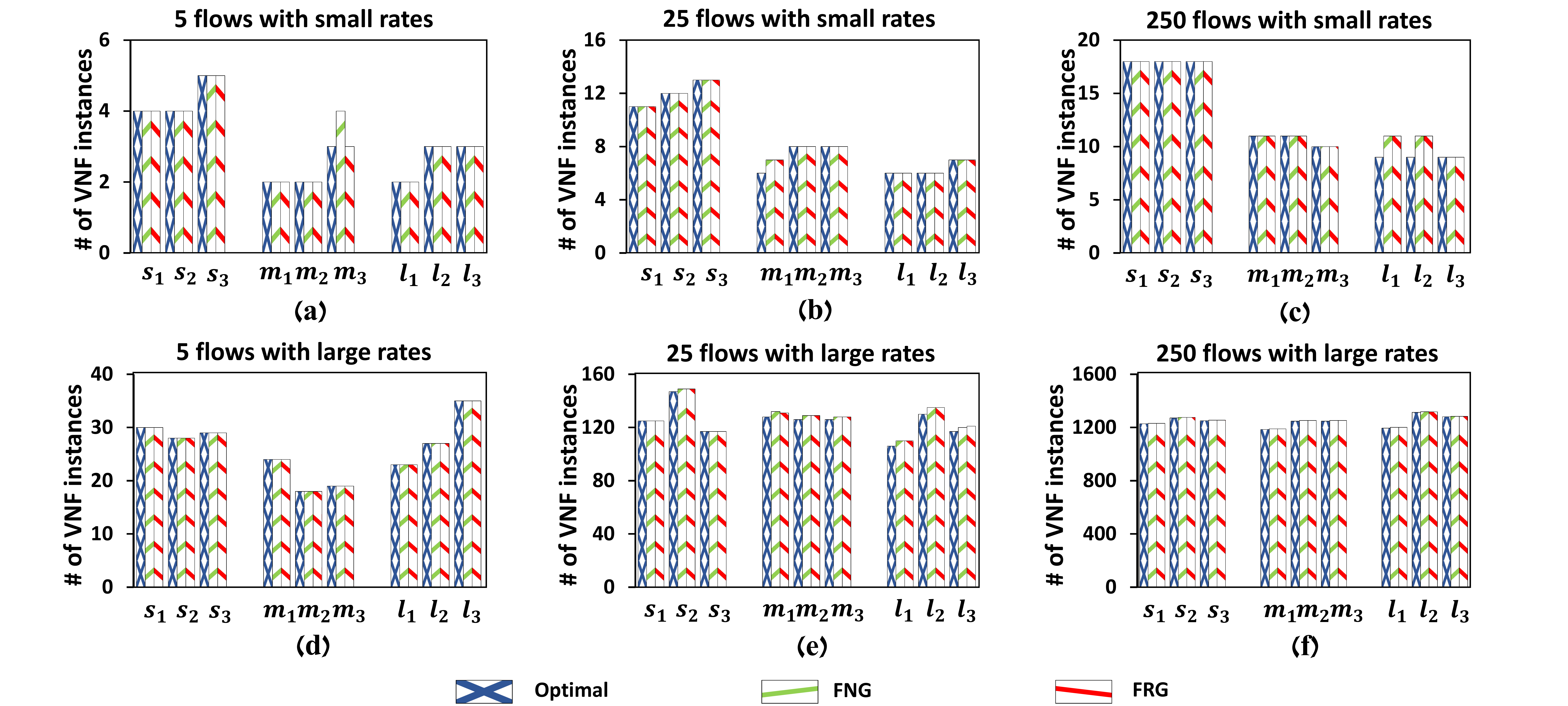}
				\caption{Simulation results for InternetMCI.}
				\label{fig:simulation25}
\end{figure*}

In order to obtain a comprehensive understanding of the empirical performance of our algorithms, we conduct simulations in various scenarios.
Specifically, we consider the following settings: (i) different path lengths, (ii) different flow rates, and (iii) different number of flows. 
The simulation results are shown in Fig.~\ref{fig:simulation40} and Fig.~\ref{fig:simulation25}. 
Each set of simulation results consists of 6 subfigures; each subfigure consists of the results of three different settings; each setting has three different simulation instances.
We use the title and the $x$ axis to distinguish different simulation settings.
Along the $x$ axis, we use ``s", ``m", and ``l" to denote the setting where flows have short paths, medium paths, and long paths, respectively.
Label ``$s_i$" denotes the $i$-th simulation instance of the short-path setting.
Other labels have similar meanings.
The y axis is the total number of VNF instances used in the network.
We will discuss the impact of path length, flow rate, and topology complexity.
In the last part, we point our another nice property of our greedy algorithms that is not  explicitly mentioned in previous sections.
The detailed simulation results are presented in the following.

\textbf{Impact of path length:}
We consider the following ranges for each type of flow path. 
A short path has a length uniformly distributed in the range of $[1,n/10]$ hops, where $n$ is the number of nodes in the network. 
Similarly, the range for the length of a medium path and of a long path is $[1, n/4]$ hops and $[1, n/2]$ hops, respectively. 
A larger average path length implies that the flows have a bigger chance to intersect each others.
This provides a larger room for optimization by combining the processing of multiple flows at fewer nodes.
Therefore, the total number of VNF instances would decrease as the average path length becomes larger, especially when the flow rates are small (see Fig~\ref{fig:simulation40}-(a), (b), and (c)) since an isolated small rate flow can generate large resource waste.

\textbf{Impact of flow rate:}
Fig.~\ref{fig:simulation40}-(a), (b), and (c) show the results where all the flows have small rates. 
The flow rates are uniformly distributed in the range of $\left[0, R/m\right]$, where $m$ is the number of flows. 
Fig.~\ref{fig:simulation40}-(d), (e), and (f) correspond to the results for large flow rates, which are uniformly distributed in the range of $\left[0,10R\right]$. 
In both cases, the solutions generated by the greedy algorithms are very close to the optimal solution. 
An intuitive explanation is the following.
When the flow rates are large, the density of the solution given by our algorithms is typically large (e.g., larger than 10). 
This leads to  an approximation ratio close to 1 due to Lemma~\ref{lemma:constant}.

\textbf{Impact of topology complexity: }The above simulation results show that our proposed greedy algorithms empirically perform very well in a randomly generated dense network topology. However, in reality backbone network topologies are typically sparse. To that end, we also repeat our evaluations for a realistic backbone network topology of InternetMCI (Fig.~\ref{fig:internetmci}). The simulation results are presented in Fig.~\ref{fig:simulation25}.
As shown in the plots, when the topology of a network is more sparse, the performance of our proposed greedy algorithms becomes closer to that of the optimal solution. 
The reason is the following.
When the network is smaller and the topology is more sparse, the room for optimization becomes smaller, and thus, the performance gap between our proposed greedy algorithms and the optimal solution also reduces.

\begin{figure}[!t]
		\centering
        \includegraphics[width=0.4\textwidth]{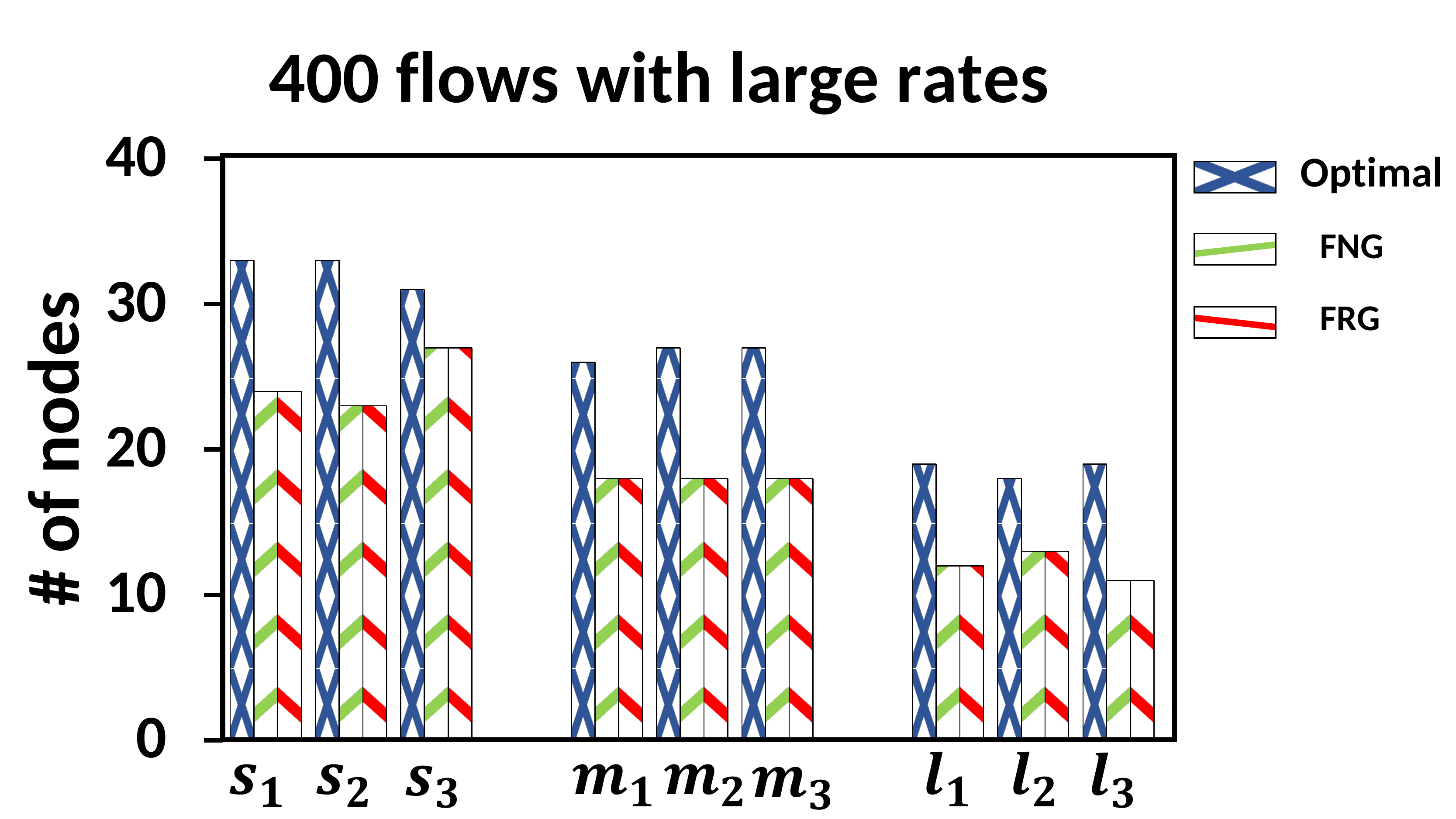}
        \caption{The number of nodes used to host the VNF instances.}
        \label{fig:nodes}
\end{figure}
\textbf{Number of nodes used to host VNF instances: } 
In practice, the operation cost of using NFV also depends on the number of nodes that host VNF instances. 
When there is no VNF running on a physical machine, the machine can be set to sleep to reduce the energy cost.
While we did not explicitly model this cost in our objective function, the greedy nature of our algorithms guarantees that they use as few nodes as possible to host all the VNF instances.
The simulations results verify our analysis and show that our algorithms use fewer nodes to host the VNF instances to process all the flows than that used by the optimal solution in most cases we consider. 
The case of small rate is trivial as the number of VNF instances is the same as the number of nodes used to host them.
However, for the case of large rates, the solutions generated by our greedy algorithms use much fewer nodes than the optimal solution.
We show the number of nodes used to host the VNF instances under the setting of 400 flows with large rates in Fig.~\ref{fig:nodes}.
In order to achieve optimality, the optimal solution has to use almost twice as many nodes as the greedy algorithms use to host the VNF instances.
Given the complexity and limited performance improvement (less than $4\%$ according to the simulation results) of the optimal solution, we believe that the proposed greedy algorithms will be important in application scenarios of practical interest.

\section{Conclusion}\label{sec:conclusion}
In this paper, we studied the problem of joint placement and allocation of VNF instances in a new NFV-enabled networking paradigm. 
We proved that the formulated problem is NP-hard.
Then, we proposed two simple greedy algorithms that are asymptotically optimal in general topologies and an optimal greedy algorithm for tree topologies. The simulation results elucidated our theoretical analyses. 
We believe that our analytical results provide important insights that will be useful in practice. 
However, several important issues remain unaddressed.
First, we have assumed that the flow routes are fixed. 
It would be interesting to investigate the problem of joint VNF placement and flow routing. 
Second, we considered a simplified model that has only one single network function.
It would be important to account for the practical constraint of service function chaining and design new algorithms with provable performance guarantees in such settings.

\bibliographystyle{IEEEtran}
\bibliography{nfv}

% Generated by IEEEtran.bst, version: 1.14 (2015/08/26)
\begin{thebibliography}{10}
\providecommand{\url}[1]{#1}
\csname url@samestyle\endcsname
\providecommand{\newblock}{\relax}
\providecommand{\bibinfo}[2]{#2}
\providecommand{\BIBentrySTDinterwordspacing}{\spaceskip=0pt\relax}
\providecommand{\BIBentryALTinterwordstretchfactor}{4}
\providecommand{\BIBentryALTinterwordspacing}{\spaceskip=\fontdimen2\font plus
\BIBentryALTinterwordstretchfactor\fontdimen3\font minus
  \fontdimen4\font\relax}
\providecommand{\BIBforeignlanguage}[2]{{%
\expandafter\ifx\csname l@#1\endcsname\relax
\typeout{** WARNING: IEEEtran.bst: No hyphenation pattern has been}%
\typeout{** loaded for the language `#1'. Using the pattern for}%
\typeout{** the default language instead.}%
\else
\language=\csname l@#1\endcsname
\fi
#2}}
\providecommand{\BIBdecl}{\relax}
\BIBdecl

\bibitem{nfv}
\BIBentryALTinterwordspacing
``Etsi. network functions virtualisation – introductory white paper.'' 2012.
  [Online]. Available: \url{https://portal.etsi.org/NFV/NFV_White_Paper.pdf}
\BIBentrySTDinterwordspacing

\bibitem{realtime}
Y.~Li, L.~Phan, and B.~T. Loo, ``Network functions virtualization with soft
  real-time guarantees,'' in \emph{Proceedings of IEEE INFOCOM}, 2016.

\bibitem{sdn}
D.~Kreutz, F.~M. Ramos, P.~Esteves~Verissimo, C.~Esteve~Rothenberg,
  S.~Azodolmolky, and S.~Uhlig, ``Software-defined networking: A comprehensive
  survey,'' \emph{Proceedings of the IEEE}, vol. 103, no.~1, pp. 14--76, 2015.

\bibitem{du1}
A.~Olteanu, Y.~Xiao, K.~Wu, and X.~Du, ``An optimal sensor network for
  intrusion detection,'' in \emph{Proceedings of IEEE International Conference
  on Communications}, 2009, pp. 1--5.

\bibitem{du2}
J.~Lv, W.~Yang, L.~Gong, D.~Man, and X.~Du, ``Robust wlan-based indoor
  fine-grained intrusion detection,'' \emph{Proceedings of IEEE GLOBECOM},
  2016.

\bibitem{nearoptimal}
R.~Cohen, L.~Lewin-Eytan, J.~S. Naor, and D.~Raz, ``Near optimal placement of
  virtual network functions,'' in \emph{Proceedings of IEEE INFOCOM}, 2015.

\bibitem{singleservice}
M.~Casado, T.~Koponen, R.~Ramanathan, and S.~Shenker, ``Virtualizing the
  network forwarding plane,'' in \emph{Proceedings of the Workshop on
  Programmable Routers for Extensible Services of Tomorrow}, 2010, p.~8.

\bibitem{orchestra}
M.~F. Bari, S.~R. Chowdhury, R.~Ahmed, and R.~Boutaba, ``On orchestrating
  virtual network functions,'' in \emph{Proceedings of 11th International
  Conference on Network and Service Management (CNSM)}, 2015, pp. 50--56.

\bibitem{stratos}
A.~Gember, A.~Krishnamurthy, S.~S. John, R.~Grandl, X.~Gao, A.~Anand,
  T.~Benson, V.~Sekar, and A.~Akella, ``Stratos: A network-aware orchestration
  layer for virtual middleboxes in clouds,'' \emph{arXiv preprint
  arXiv:1305.0209}, 2013.

\bibitem{opennf}
A.~Gember-Jacobson, R.~Viswanathan, C.~Prakash, R.~Grandl, J.~Khalid, S.~Das,
  and A.~Akella, ``Opennf: Enabling innovation in network function control,''
  \emph{ACM SIGCOMM Computer Communication Review}, vol.~44, no.~4, pp.
  163--174, 2015.

\bibitem{mohammadkhan2015virtual}
A.~Mohammadkhan, S.~Ghapani, G.~Liu, W.~Zhang, K.~Ramakrishnan, and T.~Wood,
  ``Virtual function placement and traffic steering in flexible and dynamic
  software defined networks,'' in \emph{The 21st IEEE International Workshop on
  Local and Metropolitan Area Networks}, 2015, pp. 1--6.

\bibitem{addis2015virtual}
B.~Addis, D.~Belabed, M.~Bouet, and S.~Secci, ``Virtual network functions
  placement and routing optimization,'' in \emph{Proceedings of IEEE 4th
  International Conference on Cloud Networking (CloudNet)}, 2015.

\bibitem{kuo2016deploying}
T.-W. Kuo, B.-H. Liou, K.~C.-J. Lin, and M.-J. Tsai, ``Deploying chains of
  virtual network functions: On the relation between link and server usage,''
  in \emph{Proceedings of IEEE INFOCOM}, 2016.

\bibitem{internetmci}
\BIBentryALTinterwordspacing
 [Online]. Available: \url{http://topology-zoo.org/maps/Internetmci.jpg}
\BIBentrySTDinterwordspacing

\bibitem{setcover}
U.~Feige, ``A threshold of ln n for approximating set cover,'' \emph{Journal of
  the ACM (JACM)}, vol.~45, no.~4, pp. 634--652, 1998.

\bibitem{cloud4nfv}
J.~Soares, M.~Dias, J.~Carapinha, B.~Parreira, and S.~Sargento, ``Cloud4nfv: A
  platform for virtual network functions,'' in \emph{Proceedings of IEEE 3rd
  International Conference on Cloud Networking (CloudNet)}, 2014, pp. 288--293.

\bibitem{cloud4nfv2}
J.~Soares, C.~Goncalves, B.~Parreira, P.~Tavares, J.~Carapinha, J.~P. Barraca,
  R.~L. Aguiar, and S.~Sargento, ``Toward a telco cloud environment for service
  functions,'' \emph{Communications Magazine, IEEE}, vol.~53, no.~2, pp.
  98--106, 2015.

\bibitem{cdn}
S.~Seyyedi and B.~Akbari, ``Hybrid cdn-p2p architectures for live video
  streaming: Comparative study of connected and unconnected meshes,'' in
  \emph{Proceedings of International Symposium on Computer Networks and
  Distributed Systems (CNDS)}, 2011, pp. 175--180.

\bibitem{glpk}
\BIBentryALTinterwordspacing
 [Online]. Available: \url{https://www.gnu.org/software/glpk/}
\BIBentrySTDinterwordspacing

\end{thebibliography}

\end{document}